\documentclass[a4paper,11pt]{article}
\pdfoutput=1 

\usepackage{jcappub} 

\usepackage[T1]{fontenc} 
\usepackage{amsmath}
\usepackage{graphicx}
\usepackage{esvect}
\usepackage{multicol}
\usepackage{adjustbox}
\usepackage{slashbox,pict2e}
\usepackage{booktabs}
\usepackage[table]{xcolor}
\usepackage{array}
\usepackage[utf8]{inputenc}
\usepackage{subcaption}
\usepackage{floatrow}
\usepackage{gensymb}
\title{\boldmath Decaying Fermionic Dark Matter Search with CALET}

\author[a,1]{S. Bhattacharyya ,\note{Corresponding author.}}
\author[b]{H. Motz,}
\author[a,c]{S. Torii,}
\author[c]{Y. Asaoka}

\affiliation[a]{Advanced School for Science and Engineering,\\Waseda University, 3-4-1, Okubo, Shinjuku, Tokyo, 169-8555, Japan}
\affiliation[b]{International Center for Science and Engineering Programs,\\Waseda University, 3-4-1, Okubo, Shinjuku, Tokyo, 169-8555, Japan}
\affiliation[c]{Research Institute for Science and Engineering,\\Waseda University, 3-4-1, Okubo, Shinjuku, Tokyo, 169-8555, Japan}

\emailAdd{saptashwab@ruri.waseda.jp}
\emailAdd{motz@aoni.waseda.jp}
\emailAdd{torii.shoji@waseda.jp}
\emailAdd{yoichi.asaoka@aoni.waseda.jp}

\abstract{The ISS-based CALET (CALorimetric Electron Telescope) detector can play an important role in indirect search for Dark Matter (DM),  measuring the electron+positron flux in the TeV region for the first time directly. With its fine energy resolution of approximately $2\%$ and good proton rejection ratio ($1:10^5$) it has the potential to search for fine structures in the Cosmic Ray (CR) electron spectrum. In this context we discuss the ability of CALET to discern between signals originating from astrophysical sources and DM decay. We fit a parametrization of the local interstellar electron and positron spectra to current measurements, with either a pulsar or 3-body decay of fermionic DM as the extra source causing the positron excess. The expected CALET data for scenarios in which DM decay explains the excess are calculated and analyzed. The signal from this particular 3-body DM decay which can explain the recent measurements from the AMS$-02$ experiment is shown to be distinguishable from a single pulsar source causing the positron excess by 5 years of observation with CALET, based on the shape of the spectrum. We also study the constraints from diffuse $\gamma$-ray data on this DM-only explanation of the positron excess and show that especially for the possibly remaining parameter space a clearly identifiable signature in the CR electron spectrum exists.}

\begin{document}
\maketitle
\flushbottom

\section{Introduction}
\label{sec:intro}

While the existence and cosmological properties of Dark Matter (DM) are well established, nature and particle properties of DM are largely unknown.  Many theoretical models predict that a TeV scale Cold DM (CDM) can decay or annihilate into Standard Model (SM) particles. As a result, CDM could be detected indirectly by observing an excess in cosmic ray (CR) spectra relative to the astrophysical background~\cite{Cirelli:2008pk}. Recent results from space based CR detectors such as AMS-$02$~\cite{Aguilar:2014mma} and PAMELA~\cite{Adriani:2008zr} show an increase of the positron fraction above $10$ GeV up to $300$ GeV which is not expected from the secondary production of positrons in the Interstellar Medium (ISM).  This excess may be explained by an extra source emitting  \mbox{electron-positron} pairs, such as emission from pulsars or decay and annihilation of DM \cite{Serpico:2011wg}. To explain the positron excess with DM annihilation would require a large boost factor because the cross section of DM annihilation from relic density measurements~\cite{Adam:2015rua} yields a positron flux which is too low to produce the excess observed in the measurements~\cite{Bergstrom:2008gr,Cholis:2008hb}. The DM decay scenario can naturally explain the positron excess if the lifetime of the DM is less than~\mbox{$\sim 10^{25}$~s \cite{Cirelli:2008pk,Nardi:2008ix}}. Among different DM decay scenarios, a 3-body leptonic decay is favorable to explain the recent positron excess, because the 3-body decay produces a softer spectrum compared to \mbox{2-body} decay. Moreover, since the decay products are only leptonic, the absence of a hadronic component allows for compatibility with the recent \mbox{anti-proton measurements~\cite{Aguilar:2016kjl}.}  
\par 


In this paper, we will present the prospects of discerning such a signal from decaying DM with 1\textendash 2 TeV mass from a single pulsar source in the $(e^++e^-)$ spectrum by the measurement taken with the CALorimetric Electron Telescope (CALET). CALET, in operation on the ISS since October 2015, is designed to search for signatures from nearby CR sources and DM in this spectrum with fine energy resolution of approximately $2\%$ and high proton rejection power $(1:10^5)$ \cite{Torii:2015lck, Asaoka:2017qfm}. \par   

We study a DM candidate undergoing 3-body decay into two charged leptons and a neutrino, as a possible extra source which can explain the excess of the positron fraction observed by AMS-02~\cite{Kohri:2013sva}. 
The AMS-02 collaboration proposed an extra source emitting electron-positron pairs with an exponentially cut-off power-law spectrum~\cite{Aguilar:2013qda} as an empirical model to the positron excess. This spectrum corresponds well to that of a single young pulsar~\cite{Motz:2015cua}, making it a generic scenario against which we test the DM model explaining the positron excess. This  parametrization for the positron fraction is extended to the $(e^++e^-)$ flux and into the TeV region, including effects of propagation in the galaxy. The free parameters of this local CR parametrization with DM or Pulsar as extra source are determined from the best fit to AMS-02 positron flux and $(e^++e^-)$ measurements.  
Using this parametrization, we calculated the expected $(e^++e^-)$ spectrum for 5 years of observation with CALET for DM with a mass in the range of 1\textendash2 TeV, and investigate the possibility of discerning this particular DM decay from a generic single pulsar source.  \par
The recent diffuse $\gamma$-ray data measured by the Fermi-LAT experiment~\cite{Ackermann:2012pya} gives a strong constraint on DM annihilation or decay in the galactic halo. We compare the $\gamma$-ray emission predicted by this DM model with the $\gamma$-ray measurement and  show that $\gamma$-ray production can be reduced significantly, when the charged primary decay products from the DM are only electron and muon, excluding tau leptons. The ability of CALET to discern the DM signal from a single pulsar depends on the shape of the $(e^++e^-)$ decay spectrum, and we show this scenario with low $\gamma$-ray yield would have an especially well distinguishable signature.  

\begin{samepage}
\section{3-Body Decay of Dark Matter and the Cosmic Ray Positron Excess}\label{sec:DMmodel}
 
To explain the positron excess, various particle physics models with a 3-body decay of DM are proposed~\cite{Choi:2010jt,Cheng:2012uk,Ibe:2014qya}. In this context, we investigate a scenario where a TeV scale DM decays to leptons  $(\text{DM}\rightarrow l^-l^+ \nu)$, namely a charged standard model lepton+anti-lepton pair and a neutrino. The branching ratios of the outgoing leptons are proportional to the inverse of the decay time $(\frac{1}{\tau _e},\,\frac{1}{\tau _{\mu}},\,\frac{1}{\tau _{\tau}})$ of the DM for the individual decay channels $(ee\nu,\,\mu\mu\nu,\,\tau\tau\nu)$. We treat these as free parameters in our study and adjust them to explain the positron excess. In a recently proposed theoretical model, this type of DM decay is predicted by extending the SM with 3 fermionic singlets $N_L,\,\psi_R,\,S_R$ and two Higgs doublets $\eta,\,\chi$~\cite{Kohri:2013sva}. \par In this model, the visible matter and DM are all created from the decay of the scalar fields $(\eta ,\,\chi)$, which are charged under the $U(1)_{(B-L)}$ group and created from the decay of a generic hidden sector scalar field $\phi$. These processes occur above electro-weak scale and the predicted lifetime of the DM $( 5\times 10^{25}\,$\textendash$\, 6\times 10^{25}\,\text{s})$  is larger than the age of the universe if the $B-L$ symmetry is assumed to be broken above TeV scale, yielding the correct relic abundance. The smallness of neutrino masses and the matter-antimatter asymmetry also appear as consequences of this theoretical concept. The DM candidate is the lightest fermion $N_L$,  which decays under violation of the lepton number by two units, contributing to the CR lepton spectra. 
\par In the decaying DM scenario, the injected particles per volume and time are given by
\begin{equation}
Q = \Gamma \frac{\rho}{\text{M}_{\text{DM}}}\frac{dN}{dE}
\end{equation} 
where $\Gamma$, $\text{M}_{\text{DM}}$ are the decay rate and mass of the DM respectively.  Since the decay of the DM is mediated by a heavy scalar, the lifetime of the mediator is negligible, making 4-point scalar interaction a good approximation. With these assumptions the probability distribution for the momentum of the charged leptons $(e^+e^-,\,\mu ^+\mu ^-,\,\tau ^+\tau^ -)$ is given by 
\begin{equation}
 \label{eq:decaywidth2}
\frac{1}{\Gamma}\frac{d\Gamma}{dx}=2x ^2(3-2x)
\end{equation}  
where $x=E/E_{\text{max}}$ and $E_{\text{max}}=0.5\,\text{M}_{\text{DM}}$.  \nopagebreak 
\end{samepage}
\clearpage
From this initial energy distributions, the \mbox{$e^+$ and $e^-$} spectrum  $\frac{dN}{dE}$ produced per decay is calculated using the event generator PYTHIA (Version $8.2$)~\cite{Sjostrand:2007gs}. The spectra for $e^+$ and $e^-$ are identical and the $e^+$ spectrum is propagated in GALPROP~\cite{galprop,Strong:1998pw}. The propagation parameters in GALPROP, which is modified to include the spiral arm nature of the galaxy, are determined from comparing the background CR propagation calculation (Proton spectrum and $B/C$ ratio) with AMS-02 measurements, which is discussed in Appendix \ref{sec:GALPROPparam}. We assume a Navarro-Frenk-White (NFW) profile~\cite{Navarro:1996gj} for the DM distribution in our galaxy.
\begin{equation}
 \label{eq:NFWprofile}
\rho = \frac{\delta _c \rho _c(r/r_s)}{(1+r/r_s)^2}
\end{equation}
$\delta _c$ is defined as  
\begin{equation}
 \label{eq:nfw}
\delta _c = \frac{200}{3}\frac{c_v ^3}{\text{ln}(1+c_v)+(c_v/1+c_v)} 
\end{equation} 
where $c_v$ is defined as the ratio of virial radius $(r_v)$ and scale radius $(r_s)$, and we assume \mbox{$c_v=10$~\cite{Lokas:2000mu}}. $\rho _c$ is determined from the mass of the halo as 
\begin{equation}
\rho _c = \frac{\frac{4}{3}\pi r_v ^3}{M_v} 
\end{equation}
where $r_v,\,M_v$ are taken as $200$ kpc and $1.5\times 10^{12}\text{M}_{\odot}$~\cite{Dehnen:2006cm}.

\section{Parametrization of Local $e^+$ and $e^-$ Flux and Fit to Current Data}\label{sec:Para}
The locally observed $e^+$ and $e^-$ spectra are parametrized to reflect the variability from the free parameters of injection and propagation. Using this parametrization we determine multiple scenarios for DM as the extra source explaining the positron excess from the minimum $\chi ^2$ in comparison with the $(e^++e^-)$ and positron flux measurements from AMS-02~\cite{Aguilar:2014mma}. The parametrization is based on the assumption that distant supernova remnants (SNR) give a power law primary electron spectrum, to which a secondary component from nuclei interactions with the ISM is added. We also assume that the injection spectrum index of electrons and nuclei is the same as they originate from the same sources. This is described by two power law indices $\gamma _p,\, \gamma _{s}$ and two coefficients $C_p,\,C_{s}$  which describe the relative weights of the spectra for primary electron and secondary flux. The radiative energy loss processes (such as synchrotron radiation, Inverse Compton radiation, Coulomb scattering etc.) experienced by the primary electrons are modeled as an exponential cut-off at energy $E_d$, which is absent for the secondary particles. With these parameters the total flux (primary+secondary) can be written as
\begin{equation}
 \label{eq:totpara}
\phi _T(E)= 2\phi _{\text{extra}}+C_pE^{\gamma _p}\left(2\frac{C_{s}}{C_p}E^{\gamma _{s} -\gamma _p}+e^{(\frac{-E}{E_d})}\right) 
\end{equation}
where $\phi _{\text{extra}}$ is the flux from  the extra sources emitting electron-positron pairs. \clearpage 
For the pulsar scenario we parametrize the extra source by
\begin{equation}
 \label{eq:extrasourcepulsar}
\phi _{pn} = C_{pn}E^{\gamma _{pn}}e^{-\left(\frac{E}{E_{pn}}\right)} 
\end{equation}
here the weight of the diffuse spectra is given by $C_{pn}$, power law index $\gamma _{pn} $ (common for electron and positron) and a cut-off energy $E_{pn}$. \par
The extra source flux from DM decay is given by 
\begin{equation}
\phi _{\text{DM}} = \frac{1}{\tau _e}\phi _e + \frac{1}{\tau _{\mu}}\phi _{\mu} + \frac{1}{\tau _{\tau}}\phi _{\tau} 
\end{equation}
with $\phi _e,\,\phi _{\mu},\,\phi _{\tau}$ being the $e^+$ (identical to $e^-$) decay spectra for $ee\nu,\,\mu\mu\nu,\,\tau\tau\nu$ channel respectively, propagated with GALPROP and $\frac{1}{\tau _e},\,\frac{1}{\tau _{\mu}},\,\frac{1}{\tau _{\tau}}$ are the inverse of the decay times for three leptonic decay channels.   

The positron flux from eq. (\ref{eq:totpara}) can be written as
\begin{equation}
 \label{eq:posipara}
\phi _{e^+} =\phi _{\text{extra}}+C_sE^{\gamma _s}
\end{equation} 
This parametrization is fitted to the current measurements of the electron and positron flux to determine values for the free parameters.



 In this fitting $\frac{1}{\tau _e},\,\frac{1}{\tau _{\mu}},\,\frac{1}{\tau _{\tau}}$ are treated as free parameters for the DM extra source in addition to the three free parameters $C_p,\,\frac{C_s}{C_p},\,\gamma _p$ for the background (eq.~(\ref{eq:totpara})). Assuming a common origin for nuclei and electrons, the difference between the primary and secondary electron indices $(\gamma _s -\gamma _p)$ is nearly equal to $-\delta$ and thus fixed to $-0.4$ in the fit, according to the propagation model given in Appendix~\ref{sec:GALPROPparam}. The range of data points used for comparison with experimental results is from 15 GeV to 1 TeV. Since the CR spectra below 15 GeV are influenced by solar modulation, diffusive reacceleration and possibly a change in the injection index~\cite{Korsmeier:2016kha}, the variability of the spectra cannot be represented by a simple parametrization. However, we apply the effect of charge independent solar modulation above 15 GeV~\cite{Cholis:2015gna} in the parametrization by assuming force field approximation with a fixed value of 500 MeV for the common $(e^+$ and $e^-)$ modulation potential. The upper bound of the fit range is effectively 1~TeV as there are no high resolution data points  from the AMS-02 measurements above 1~TeV. The  cut-off energy $E_d$, which has only influence in the TeV region, cannot be determined from current experimental data and various values of $E_d$ (1~TeV, 2~TeV, 5~TeV, 10~TeV) are studied. To estimate the unknown $(e^++e^-)$ spectrum in the TeV region an electron-only flux from the Vela SNR, which is the most influential nearby source with distance around $1$ kpc and age less than $10^5$ years~\cite{vela:nature}, is calculated with GALPROP for the propagation parameters as described in Appendix~\ref{sec:GALPROPparam}. The contribution of Vela to the high energy electron spectrum may be reduced if the release of CR electrons is gradual or delayed. The parametrization reflects the variability of the contribution of Vela and also the influence of spiral arm thickness on the CR $(e^++e^-)$ spectrum (\ref{sec:GALPROPparam}) by choosing different values for $E_d$ in the range from 1-10~TeV. It should also be noted that a harder injection spectrum~\cite{Kobayashi:2003kp} and/or a specific energy-dependent release~\cite{Kawanaka:2010uj} of the electrons from Vela could create a distinct signature in the TeV region. If such a signature is found by CALET, the background model for DM search would have to be adapted.
\par
\begin{figure}[h]
 \includegraphics[width=1.0\textwidth]{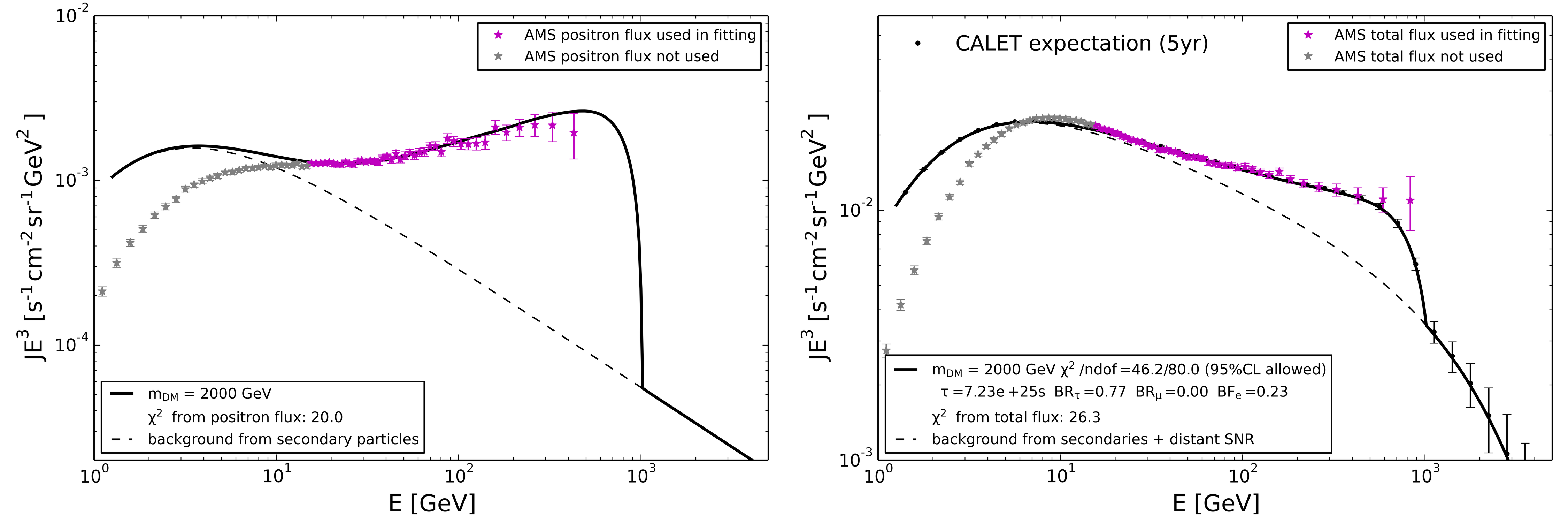}
  \caption{2 TeV fermionic DM decay spectra on top of the background (dotted line) are fitted to the AMS-02 positron flux (left panel) and $(e^++e^-)$ flux (right panel), resulting in a branching fraction of $0.77$ for $\tau\tau\nu$ channel and $0.23$ for $ee\nu$ channel (solid line). Background cut-off energy, $E_d$ is 2 TeV. \label{fig:DMasextrasource}  }
\end{figure}

 As an example we show in figure~\ref{fig:DMasextrasource} that, the fit converges at branching ratios of $0.77$ for $\tau\tau\nu$ channel and $0.23$ for $ee\nu$ channel, with no contribution from $\mu\mu\nu$ channel for a \mbox{2 TeV} fermionic DM and the background cut-off energy $(E_d)$ set to 2 TeV. \par
 
\begin{figure}[h]
 \includegraphics[width=1.0\textwidth]{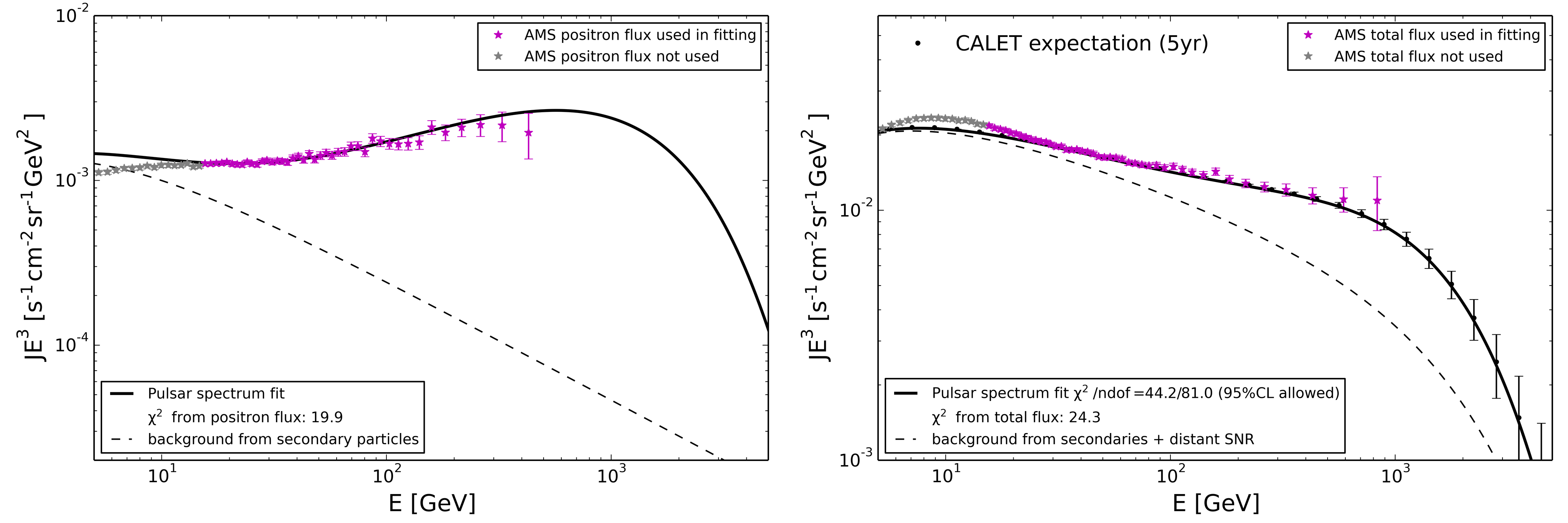}
  \caption{The parametrization of the background and a single pulsar as extra source (solid line) is fitted to the AMS-02 positron flux (left) and $(e^++e^-)$ flux (right), assuming cut-off energies $E_{pn} = 1$ TeV for the pulsar and $E_d = 2$ TeV for the background. \label{fig:Pulsarasextrasource} }
\end{figure}

Similarly, the scenario with single pulsar as the only extra source gives a good fit to
the positron flux and $(e^++e^-)$ flux in the same fit range as for DM (15 GeV - 1 TeV), shown in figure~\ref{fig:Pulsarasextrasource}. Apart from the three free background parameters,  the free parameters for a pulsar as extra source are $\frac{C_{pn}}{C_p},\, \gamma _{pn}-\gamma _p$. The values of the extra source (pulsar) free parameters are determined from the best fit assuming $E_{pn} = 1$ TeV. Since the expected CALET  data for 5~years of measurement is calculated for the DM case, the initially assumed energy cut-off for the pulsar source $(E_{pn})$ has no influence in this study, as finally when CALET's capability to discern pulsar and DM is calculated, it is taken as a free parameter. \par 
It is shown in a recent work~\cite{Feng:2015uta} that there are several candidates among pulsars within a distance of $<0.5$ kpc from the solar system and with an age of $4\times 10^4 - 4.5\times 10^5$ years which could provide a single source explanation of the positron excess. So the single young pulsar is taken as a generic case against which we compare the DM decay model.

\section{Diffuse $\gamma$-ray Constraints and Low $\gamma$-ray Flux Scenario} 

The decay or the annihilation of DM directly produces $\gamma$-rays in the from of Final State Radiation~(FSR) and also secondary $\gamma$-rays from Inverse Compton and Bremsstrahlung processes during propagation of charged decay or annihilation products. Through these processes it is expected that the decay of the investigated DM into charged leptons $(e^{\pm},\,\mu ^{\pm},\,\tau ^{\pm})$  in the galactic DM-halo would produce a  diffuse $\gamma$-ray flux. For DM decay which can explain the positron excess, this predicted $\gamma$-ray flux has to be compared with the \mbox{Fermi-LAT~\cite{Ackermann:2012pya}} diffuse $\gamma$-ray measurement taken at high latitudes. Looking away from the galactic plane $(\left|b\right|>20\degree)$ strongly reduces the background from galactic astrophysical sources and thus comparison of $\gamma$-ray flux from DM with the measurement in this region gives the strongest constraint. The remaining contribution from astrophysical sources depends on the different modelings of $\gamma$-ray emission~\cite{DiMauro:2015tfa,Ackermann:2014usa}, but the total measured flux can be considered a conservative upper bound. While the diffuse $\gamma$-ray spectrum in the relevant sky region and energy range is currently only available from Fermi-Lat, it is going to be reaffirmed by the currently operating detectors with calorimeters capable of absorbing the full shower energy up to the TeV region, such as CALET~\cite{Cannady:2015kbz} and also DAMPE~\cite{changjin}. \par 

The $\gamma$-ray flux from DM decay depends on both the mass of the decaying DM and the decay products. As the $\tau\tau\nu$ channel produces more $\gamma$-rays compared to $ee\nu$ and $\mu\mu\nu$ channel, to study the possibility of a DM-only explanation of the positron excess compatible with the current $\gamma$-ray measurements, we reduce the tau component from the decay products of the DM. Adapting all other free parameters in each step and starting with the parameters obtained from the initial fit, we reduce the tau component in steps until the $\chi ^2$ either positron flux or $(e^++e^-)$ flux exceeds $95\%$ CL, or the scale factor for $\tau\tau\nu$ channel reaches zero. The branching ratios for the initial fit and the fit with the reduced tau contribution are given in table~\ref{DMbranch} for different values of DM mass and cut-off energy $E_d$. It is shown that a good fit with completely removed $\tau\tau\nu$ channel is possible for DM with mass 1.5 TeV and 1.0 TeV, and a cut-off energy $E_d$ equal to or larger than 2 TeV or 10 TeV respectively. However, no good fit even including $\tau\tau\nu$ channel is possible for 1 TeV DM and $E_d$ equal to or smaller than 2 TeV. The chosen DM theory supports full variability of the branching fractions, which are proportional to the effective 4-point couplings for each decay mode. The effective couplings are governed by the products of the coupling constants at both vertices of the decay process which are different for each channel, making them completely free parameters also independent of the leptonic mass hierarchy~\cite{Kohri:2013sva}.       
\begin{figure}[h]
 \centering
 \includegraphics[width=1.0\textwidth]{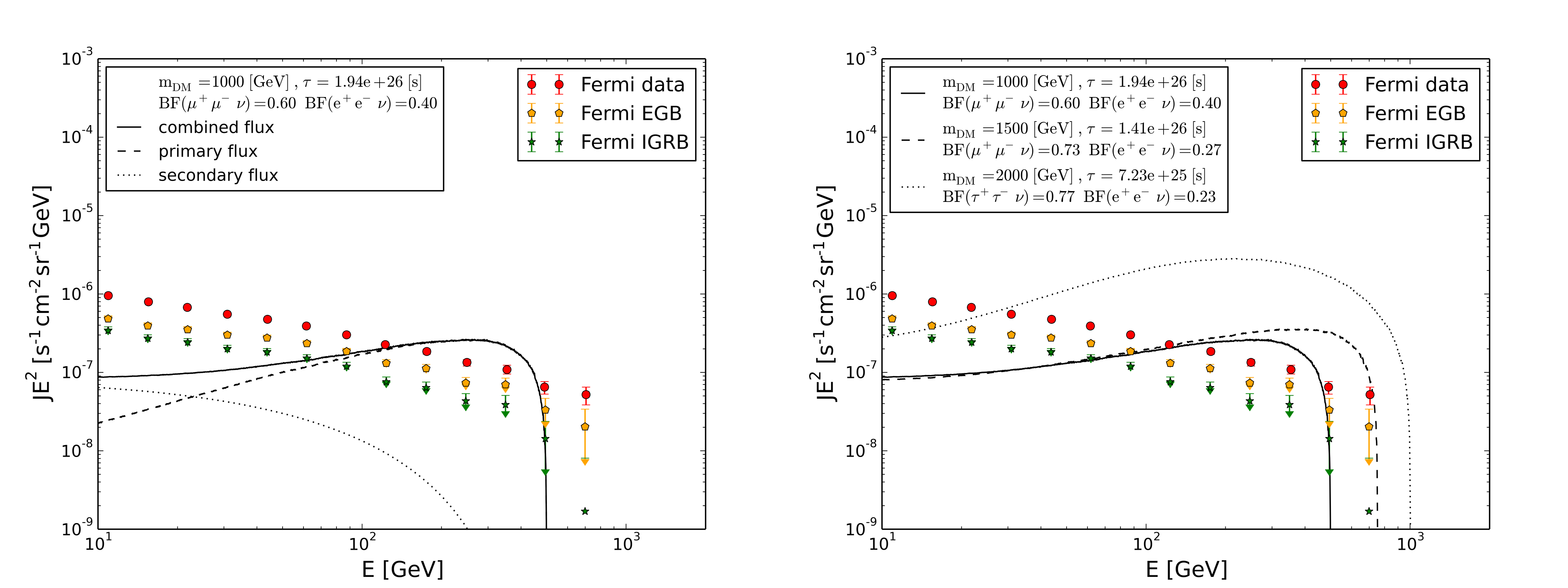}
 \caption{Predicted $\gamma$-ray flux from DM decay compared to Fermi-LAT diffuse $\gamma$-ray measurement. In the left panel we show the $\gamma$-ray flux from 1 TeV DM of the primary (Black dashed line) and secondary production (black dotted line). On the right panel the combined \mbox{$\gamma$-ray} flux from primary and secondary production are shown for 2 TeV DM $(77\%\,\tau\tau\nu ,\,23\% \,ee\nu)$ with black dotted line, 1.5~TeV DM $(73\%\,\mu\mu\nu,\,27\%\, ee\nu)$ with black dashed line, 1~TeV DM $(60\%\,\mu\mu\nu ,\,40\% \, ee\nu)$ with black solid line. \label{fig:gammalimit}}
\end{figure}

The $\gamma$-ray fluxes from the FSR and decay of the primary decay products have been calculated with PYTHIA assuming NFW profile, and three different cases are plotted in figure~\ref{fig:gammalimit} including contribution from secondary $\gamma$-rays. The charged particles from the decay of DM and their interaction with the interstellar radiation field (ISRF) produce secondary \mbox{$\gamma$-rays}. This isotropic diffuse $\gamma$-ray flux is calculated in GALPROP at latitudes $\left|b\right|> 20\degree$, for different DM models using the default ISRF~\cite{Porter:2005qx} provided by GALPROP. As shown in the left panel of figure~\ref{fig:gammalimit}, $\gamma$-rays from secondary production have lower energy than the primary component. For a DM of mass $2$ TeV decaying to $\tau\tau\nu$ $(73\%)$ and $ee\nu$ $(27\%)$ channel, the predicted $\gamma$-ray flux exceeds the the Fermi-LAT data significantly. However with $1.5$~TeV and 1~TeV DM decaying only to $\mu\mu\nu$ and $ee\nu$, the $\gamma$-ray fluxes from the decay are closer to the experimental data as shown in the right panel of figure~\ref{fig:gammalimit}. \par
\begin{figure}[h]
\centering
	\begin{subfigure}[b]{1.0\textwidth}
  	\includegraphics[width=0.97\linewidth]{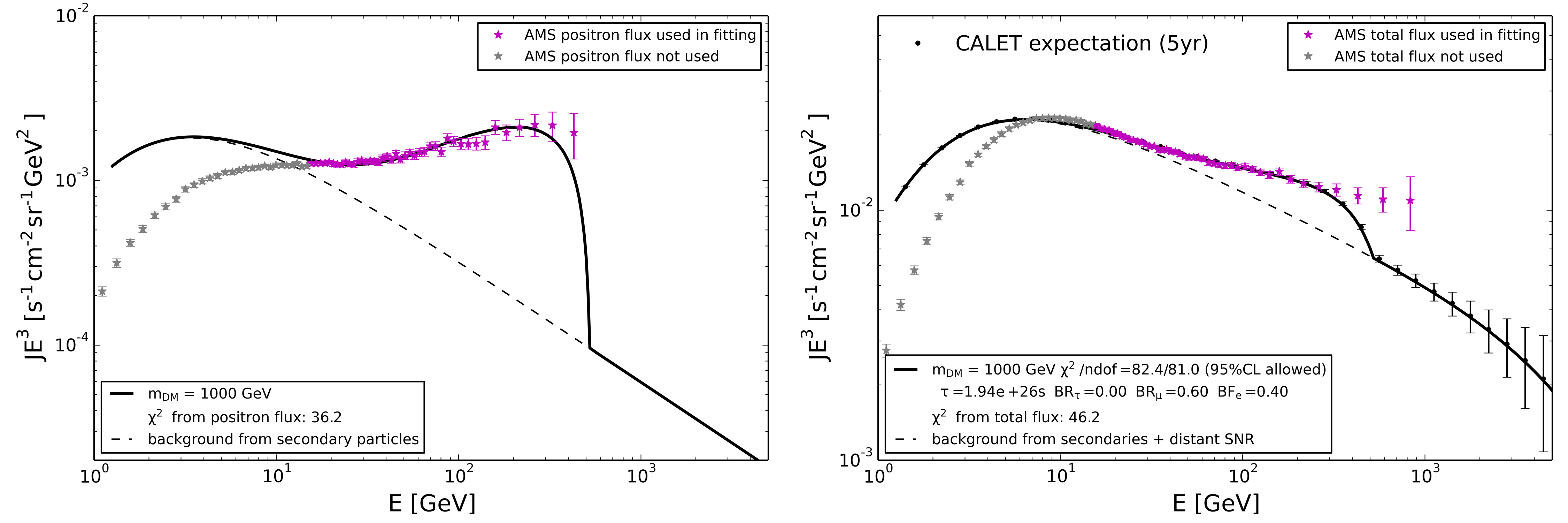}
  	\caption{\label{fig:DMfitlowgamma1}}
\end{subfigure}
	\begin{subfigure}[b]{1.0\textwidth}
  	\includegraphics[width=0.97\linewidth]{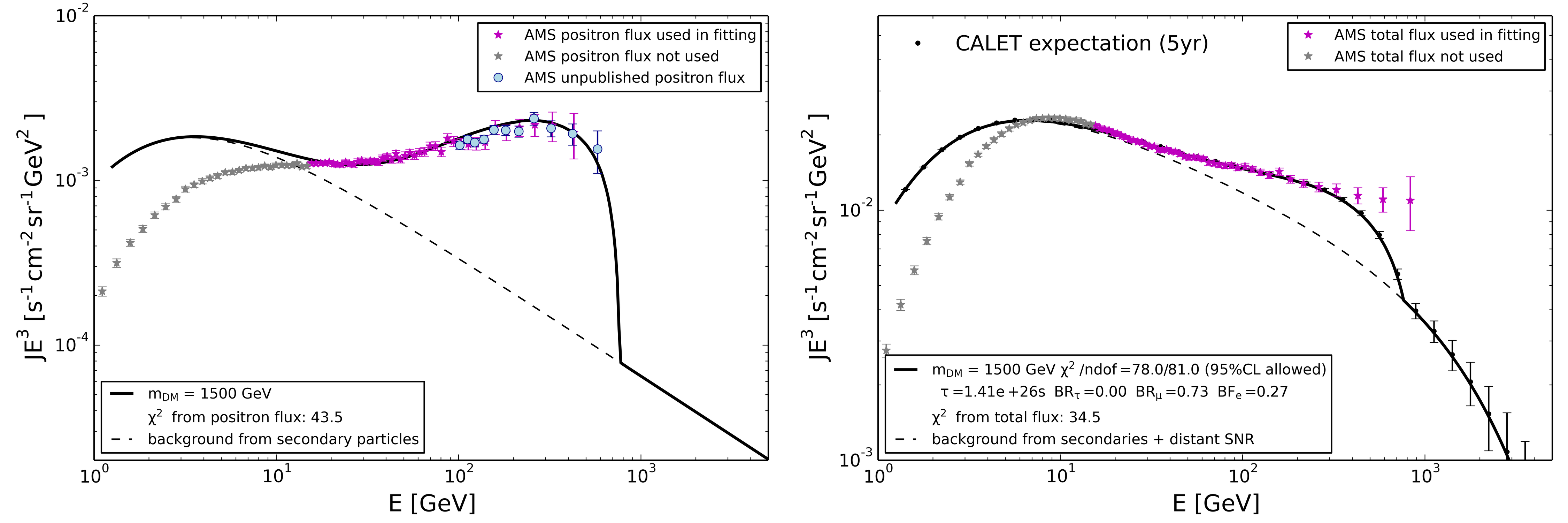}
  	\caption{\label{fig:DMfitlowgamma2}}
\end{subfigure}
\caption{(a) $1$ TeV DM (without $\tau\tau\nu$ channel) decay spectra on top of the background (dotted line) are fitted to the positron flux and $(e^-+e^+)$ flux from AMS-02. The background cut-off energy $(E_d)$ is 10 TeV. (b) As figure~\ref{fig:DMfitlowgamma1} but for a $1.5$ TeV DM with $E_d$ set to 2~TeV. This decaying DM matches well the new 5-year AMS-02 positron flux data (shown with cyan dots) which was not used in the fit.}
\end{figure}

 
The $\gamma$-ray flux from the 1 TeV DM decay scenario, as shown in figure~\ref{fig:gammalimit}, is least in conflict with the experimental data. Models with these characteristics (low DM mass, and no decay to $\tau\tau\nu$ channel) may be a unique possibility to explain the positron excess by DM, without violating the constraints from $\gamma$-ray measurements, making this model of special interest to study. For 1~TeV DM decaying only to $\mu\mu\nu$ and $ee\nu$ channel, the fit converges at branching ratios of $0.60$ for $\mu\mu\nu$ channel and $0.40$ for $ee\nu$ channel with $E_d$ set to 10~TeV as shown in figure~\ref{fig:DMfitlowgamma1}. Similarly, for a $1.5$~TeV DM decaying only to $\mu\mu\nu$ and $ee\nu$ channel the best fit converges at branching ratios of $0.73$ for $\mu\mu\nu$ channel and $0.27$ for $ee\nu$ channel  with 2~TeV background cut-off, shown in figure~\ref{fig:DMfitlowgamma2}. This $1.5$~TeV fermionic DM matches best the new AMS-02 positron flux recently presented at CERN~\cite{Ting-AMS-02}, making this another case to be studied. Although the predicted $\gamma$-ray flux from the 1~TeV DM is somewhat higher than the Fermi-LAT measurement, there should be an uncertainty in the lifetime of the DM, and thus the $\gamma$-ray flux, from the choice of propagation conditions used for the positrons of the DM decay. Also the shape of the DM halo may influence the charged CR $(e^+,\, e^-)$ and $\gamma$-ray flux. The $\gamma$-ray flux measured at higher latitudes may be reduced and the charged CR flux enhanced if the DM accumulates close to galactic plane, as in the "Dark-Disc" model~\cite{1475-7516-2017-01-021} for partly self-interacting DM. 
 
\newcommand{\specialcell}[2][c]{%
  \begin{tabular}[#1]{@{}c@{}}#2\end{tabular}}

\begin{table}[h]
	\begin{center}
		\begin{tabular}{|c|c|c|c|c|}\hline 
		\rowcolor{white} \multicolumn{5}{|c|}{Branching Ratio of DM Decay Products from Fit to AMS-02} \\
		\rowcolor{white} \multicolumn{5}{|c|}{$\frac{(ee\nu/\mu\mu\nu/\tau\tau\nu)\rightarrow\, {e,\,\mu ,\,\tau \,\text{all free}}}{(ee\nu/\mu\mu\nu/\tau\tau\nu)\rightarrow \,{\text{min.}\, \tau,\,\text{low}\, \gamma}}$} \\		
\hline
		\backslashbox{\text{M}$_{\text{DM}}$}{$E_d$ (TeV)}
		& 1 & 2 & 5 & 10 \\\hline
		2 TeV& \specialcell{$0.27/0/0.73$  \\\hline \specialcell{$ 0.23/0.53/0.24 $}} & \specialcell{$0.23/0/0.77$\cellcolor{blue!25}\\\hline \specialcell{$ 0/0.98/0.02 $} } & \specialcell{$0.20/0/0.80$\\\hline \specialcell{$ 0.09/0.83/0.08 $}} & \specialcell{$0.20/0/0.80$\\\hline \specialcell{$ 0.07/0.86/0.07 $}} \\\hline
		1.5 TeV & \specialcell{$0.30/0/0.70$\\\hline\specialcell{$ 0.34/0.59/0.07$} } & \specialcell{$0.30/0/0.70$  \\\hline \cellcolor{blue!25}\specialcell{$0.27/0.73/0$}} & \specialcell{$0.26/0/0.74$\\\hline\specialcell{$0.22/0.78/0$}} & \specialcell{$0.23/0/0.77$\\\hline\specialcell{$0.21/0.79/0$}} \\\hline
		1 TeV &Excluded&Excluded&  \specialcell{$0.27/0/0.73$\\\hline\specialcell{$0.32/0.15/0.53$}} & \specialcell{$0.26/0/0.74$ \\\hline \cellcolor{blue!25} \specialcell{$0.40/0.60/0$}} \\\hline
		\end{tabular}			
	\end{center}
	\caption{Branching ratios for fit to AMS-02 $(e^++e^-)$ and positron flux for all studied values of \text{M}$_{\text{DM}}$ and $E_d$. Upper one: initial fit, lower one: fit with $\tau\tau\nu$ component reduced. Colored boxes correspond to the examples shown in figure~\ref{fig:DMasextrasource}, figure~\ref{fig:DMfitlowgamma2} and figure~\ref{fig:DMfitlowgamma1} respectively.}
\label{DMbranch}
\end{table}

\clearpage

\section{Discerning Single Pulsar Source and Dark Matter with CALET }

\begin{figure}[h]
  \centering
  \includegraphics[width=1.0\textwidth]{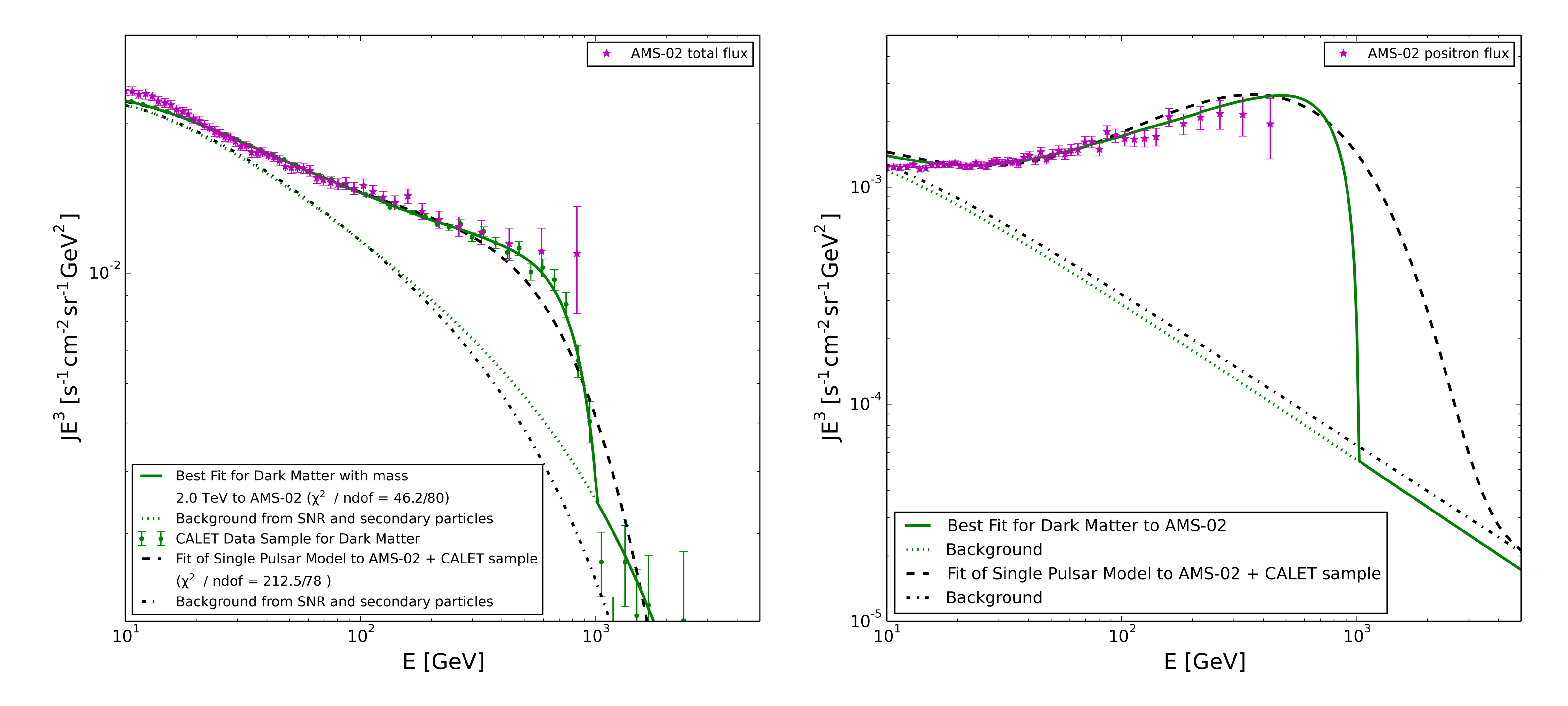}
  \caption{Fit of the single pulsar source to one of the simulated 10000 statistical samples of 5-year CALET data for $(e^-+e^+)$ flux for the 2 TeV DM (green line) and positron flux (right panel) from AMS-02 data is shown here with the black dashed lines. Background CR spectra are shown as dotted lines (green and black) for the two different extra source scenarios (DM and pulsar respectively). \label{fig:fitPulsarDM} }

\end{figure}

\begin{figure}[h]
\centering
	\begin{subfigure}[b]{1.0\textwidth}
  \includegraphics[width=1.0\linewidth]{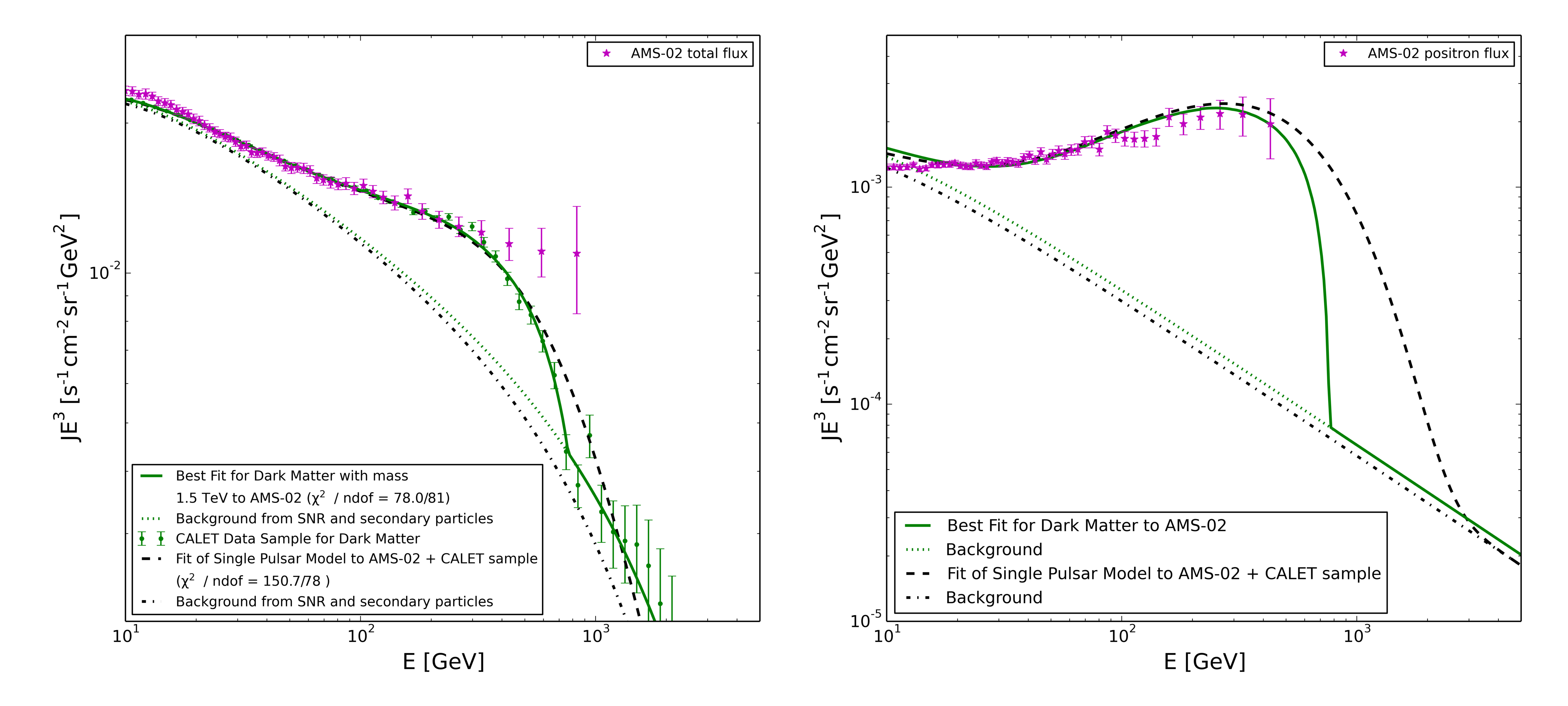}
  \caption{\label{fig:fitPulsarDMlowGamma2}}
 
\end{subfigure}

	\begin{subfigure}[b]{1.0\textwidth}
  	\includegraphics[width=1.0\textwidth]{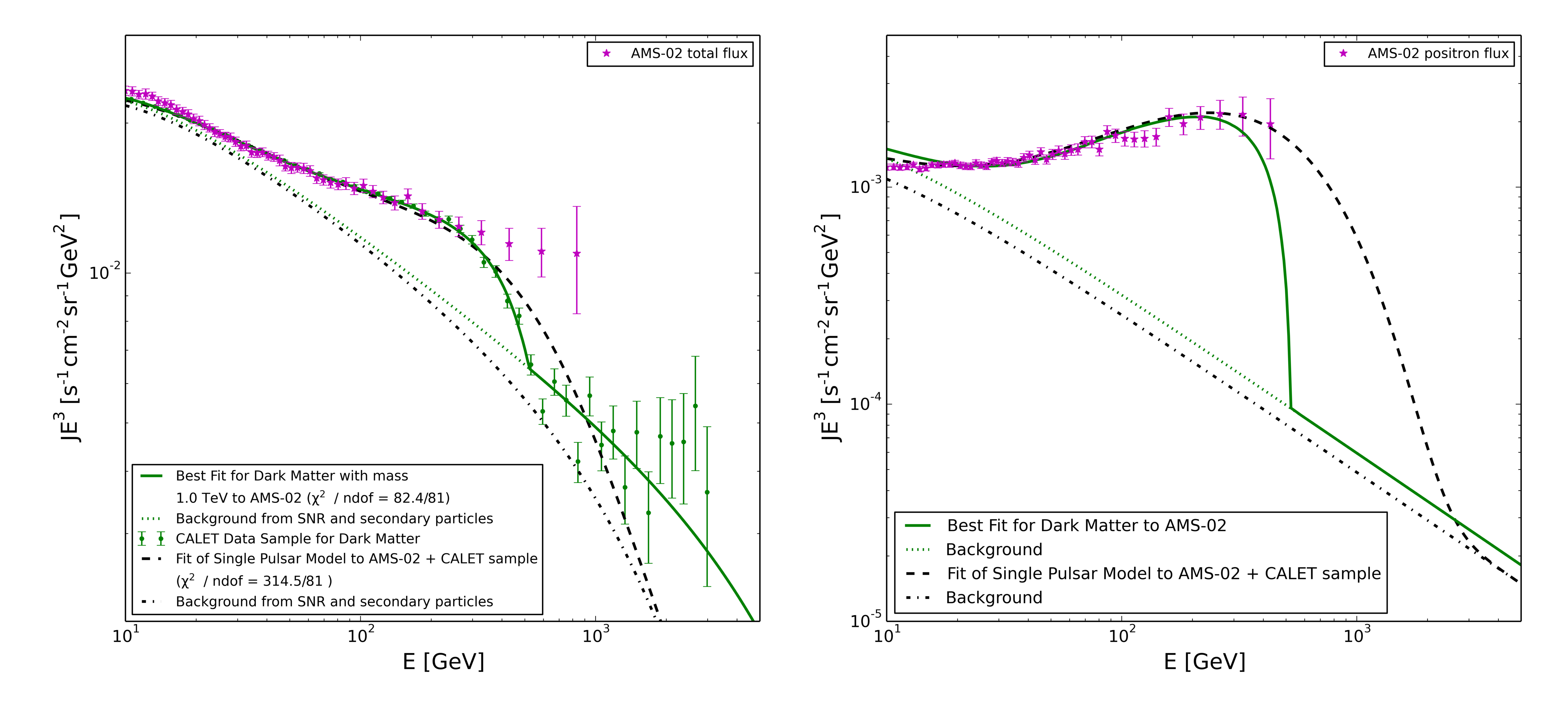}
  \caption{\label{fig:fitPulsarDMlowGamma1}}
\end{subfigure}
\caption{(a)As figure~\ref{fig:fitPulsarDM} but for $1.5$~TeV DM without $\tau\tau\nu$ channel. The background cut-off energy $(E_d)$ is 2~TeV. (b) As figure~\ref{fig:fitPulsarDMlowGamma2} for 1~TeV DM without $\tau\tau\nu$ channel with $E_d$ set to 10~TeV. }

\end{figure}

For  the fits of the parametrization with the DM source to the current experimental results as described in section \ref{sec:Para}, the expected CALET data was calculated, based on the detector's aperture of $1200\,\text{cm}^2\,\text{sr}$~\cite{akaike-ICRC} and 5 years of data-taking with a reconstruction efficiency of $90\%$. To simulate the statistical fluctuations in the event rates, $10000$ event samples were generated, representing different outcomes of the $\left(e^++e^-\right)$ flux measurement in each of the DM decay scenarios. The energy spectrum from one of these samples is shown in figure~\ref{fig:fitPulsarDM}. To find CALET's capability of discerning such a DM sample from the single pulsar source, the single pulsar source parametrization was fitted to the simulated 5-year $(e^++e^-)$ flux  CALET data for the DM   and the positron flux measured by AMS-02. In this fitting all parameters for the background parametrization $\left(C_p,\,\frac{C_s}{C_p},\,\gamma _p,\,E_d\right)$  except $\left(\gamma _s -\gamma _p\right)$ are free parameters as well as the parameters for the single pulsar source $\left(\frac{C_{pn}}{C_p},\, \gamma _{pn} -\gamma _p,\,E_{pn}\right)$. The equivalent fits of the single pulsar parametrization to a DM case sample for $1.5$ TeV and 1 TeV DM decaying only to $\mu\mu\nu$ and $ee\nu$ are shown in figure~\ref{fig:fitPulsarDMlowGamma2} and figure~\ref{fig:fitPulsarDMlowGamma1} respectively. 

\begin{figure}[h]
 \includegraphics[width=1.0\textwidth]{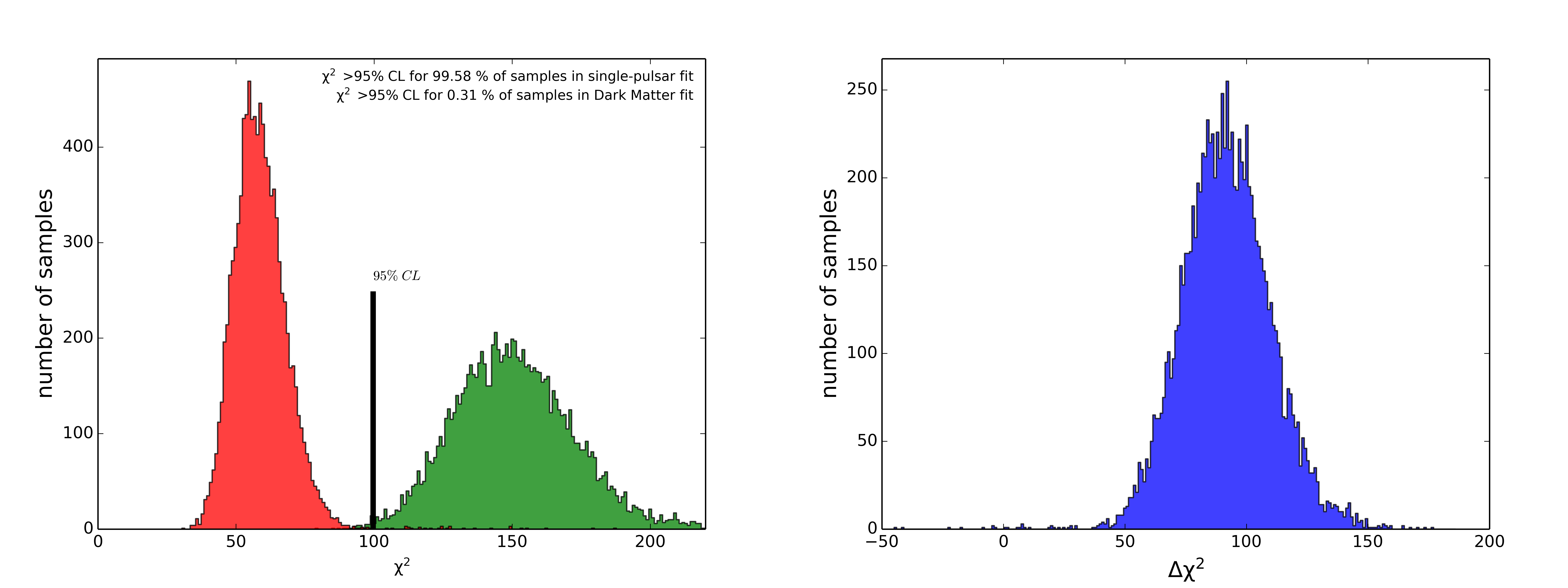}
  \caption{$\chi ^2$ distribution for the fit of the single pulsar source to the simulated CALET data for 10000 DM samples + AMS-02 positron flux data (green) and re-fit of DM samples using the same data points (red). On the right panel the difference between the $\chi ^2$ for pulsar and DM $(\chi ^2 _{\text{pulsar}}-\chi ^2 _{\text{DM}})$ is shown (blue). \label{fig:Discern}}
  
\end{figure}

We obtain the $\chi ^2$ distribution which is shown in the left panel of figure~\ref{fig:Discern} from fitting the single pulsar source to the 5-year CALET data for the 10000 simulated samples of a 2~TeV DM (for the $(e^++e^-)$ flux) and positron flux from the AMS-02 measurement. This is compared with the $\chi ^2$ distribution from re-fitting the DM model to these same data points. Since the DM and the single pulsar source model are independent of each other (non-nested), a quantitative separation strength between them such as a likelihood-ratio test statistic cannot be determined. However we can assess the quality of DM and single pulsar model relative to each other by a qualitative measure, such as Akaike's Information Criterion (AIC)~\cite{Akaikecri}, to select one model over another. The AIC value of a particular model is given by 
\begin{equation}
\text{AIC} = -2L_m + m
\end{equation}
where $L_m$ is the maximum value of the log-likelihood function and $m$ is the number of free parameters in the model. Given a set of models, the model with lowest AIC value is most favorable for representing data under the condition that the likelihood for both models follows a normal distribution. Both the pulsar model and DM model show a normal distribution which can be concluded from the $\chi ^2$ distribution plots (e.g. left panel of figure~\ref{fig:Discern}). From the definition of the single pulsar source parametrization (eq. (\ref{eq:extrasourcepulsar})), the three free parameters are the ratio of extra source coefficient to primary electron flux coefficient $\left(\frac{C_{pn}}{C_p}\right)$, the  difference of extra source power-law index to primary electron flux power-law index $(\gamma _{pn}-\gamma _p)$, and the extra source exponential cut-off energy $(E_{pn})$. For the DM model there are three free parameters, the scale factors for the three decay modes. Both cases share the four free parameters for the CR background spectra $C_p,\,\frac{C_p}{C_s},\,\gamma _p,\, E_d$ (eq. (\ref{eq:totpara})). Since both models have the same number of free parameters and the $\chi ^2$ distribution for each model resembles a normal distribution, comparison of $\chi ^2$ is equivalent to a comparison of the AIC value. As shown in the right panel of figure~\ref{fig:Discern}, the $\chi ^2$ difference $(\chi ^2 _{\text{pulsar}}-\chi ^2 _{\text{DM}})$ between single pulsar source fit and the DM re-fit is always positive except for very few samples, indicating that the simulated DM model is favored over the wrongly assumed pulsar model.  A clear discernibility can be claimed for those cases where the DM model is allowed at $95\%$ CL, while the pulsar model is excluded.  The re-fit of the DM model yields $\chi ^2 <95\%\,\text{CL}$ for all but a negligible fraction of samples as shown in table~\ref{DMchi}. Therefore, the exclusion of the pulsar case is sufficient for the separation. \par 
For the 2 TeV DM model including decay to $\tau\tau\nu$, the average $\chi ^2$ of the pulsar fit decreases with increasing $E_d$. However, still a majority of samples could be excluded even at \mbox{$E_d = 10$ TeV}, with exact numbers given in table~\ref{DMchi}. The 1.5 TeV and 1 TeV DM mass cases where no decay to $\tau\tau\nu$ takes place can be well separated from the pulsar case, independent of $E_d$.      

\begin{table}
	\begin{center}
		\begin{tabular}{|c|c|c|c|c|}\hline 
		\rowcolor{white} \multicolumn{5}{|c|}{$\frac{\text{Average}\,\chi ^2(\text{pulsar fit/DM re-fit})}{\text{No. of Samples with}\, \chi ^2\,>95\% (\text{pulsar fit/DM re-fit})}$} \\		
\hline
		\backslashbox{\text{M}$_{\text{DM}}$}{$E_d$ (TeV)}
		& 1 & 2 & 5 & 10 \\\hline
		2 TeV& \specialcell{$211.15/61.16$ \\\hline \specialcell{$10000/13$}} & \cellcolor{blue!25}\specialcell{$150.61/58.90$\\\hline \specialcell{$9958/31$} } & \specialcell{$123.87/58.31$\\\hline \specialcell{$8932/6$}} & \specialcell{$116.19/58.29$\\\hline \specialcell{$7771/1$}} \\\hline
		1.5 TeV &-& \cellcolor{blue!25}\specialcell{$142.37/75.54$  \\\hline \specialcell{$9943/138$}} & \specialcell{$132.98/76.06$\\\hline\specialcell{$9632/97$}} & \specialcell{$131.46/76.59$\\\hline\specialcell{$10000/49$}} \\\hline
		1 TeV &-&-&-& \cellcolor{blue!25}\specialcell{$269.85/72.87$ \\\hline  \specialcell{$10000/49$}} \\\hline
		\end{tabular}			
	\end{center}
	\caption{Average $\chi ^2$  obtained from the fits of the single pulsar source to the 10000 samples of simulated CALET data + positron flux from AMS-02 and the re-fit of the DM model to the same data points as a function of DM mass and background cut-off energy $(E_d)$. Colored boxes are the examples shown in figure~\ref{fig:Discern}, figure~\ref{fig:Discernlowgamma2}, figure~\ref{fig:Discernlowgamma1} respectively. In the lower panel the number of excluded samples  $(\chi ^2 >95\% \text{CL})$ for each case are shown. Average Number Degree of Freedom (NDF) is $80$ in these fits.}
\label{DMchi}
\end{table}

\begin{figure}[H]
	\begin{subfigure}[b]{1\textwidth}
	\includegraphics[width=1\linewidth]{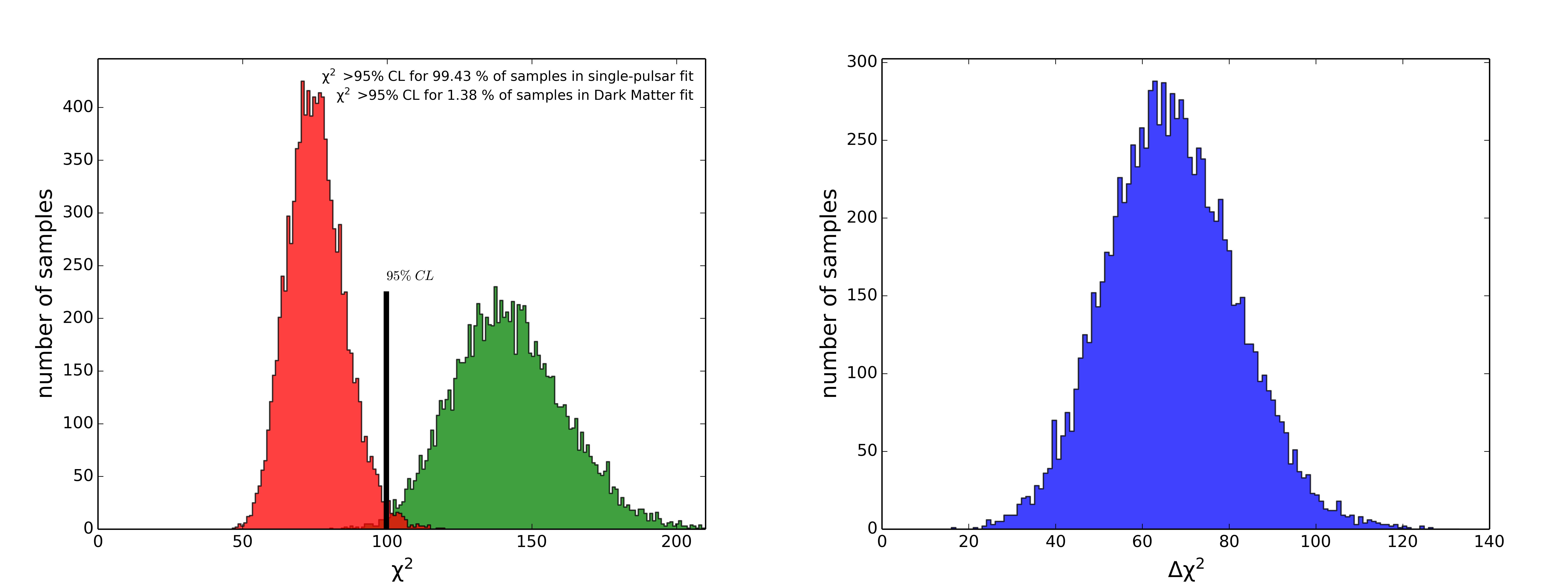}	
	\caption{\label{fig:Discernlowgamma2}}
\end{subfigure}

\begin{subfigure}[b]{1\textwidth}
	\includegraphics[width=1\linewidth]{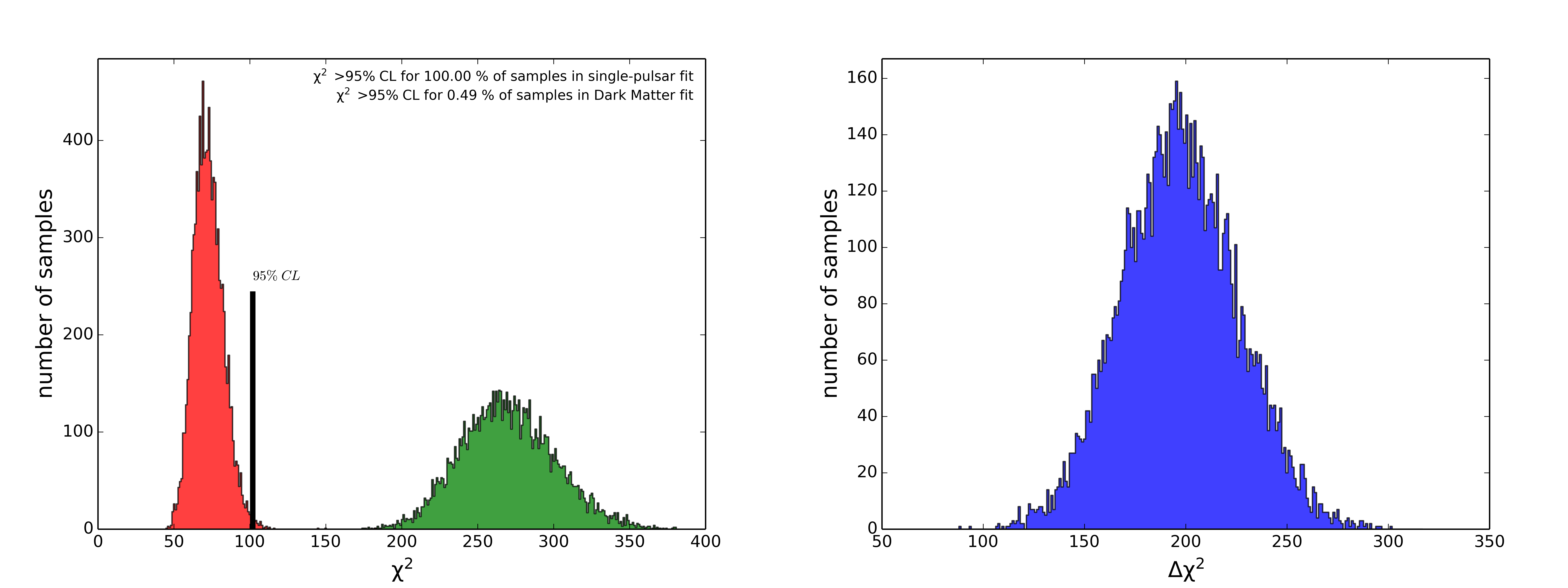}
	\caption{ \label{fig:Discernlowgamma1}}
\end{subfigure}
\caption{(a) As figure~\ref{fig:Discern}, but for the model of the $1.5$ TeV DM decay without $\tau\tau\nu$ channel with background cut-off energy $(E_d)$ set to 2 TeV. (b) As figure~\ref{fig:Discernlowgamma2}, but for the low $\gamma$-ray model of the 1 TeV DM decay without $\tau\tau\nu$ channel and $E_d$ set to 10 TeV.}
\vspace{-0.7em}	
\end{figure}

The 1 TeV DM features the largest difference between the two $\chi ^2$ distributions as shown in figure~\ref{fig:Discernlowgamma1},  demonstrating that this model of DM decay is best distinguishable from a single pulsar by the CALET $(e^++e^-)$ flux measurement. In the low $\gamma$-ray scenario with \mbox{1 TeV DM} the branching fraction obtained from the fit to the experimental results (see figure~\ref{fig:DMfitlowgamma1}) is $40\%$ for the $ee\nu$ channel. This causes a sharp drop in the $(e^++e^-)$ flux and positron flux at half the mass of the DM (see figure~\ref{fig:fitPulsarDMlowGamma1}) which is a well detectable signature. This model has the lowest predicted $\gamma$-ray flux of all the studied cases, showing a complementarity between the sensitivity of CALET and $\gamma$-ray measurements.

\section{Conclusion}

CALET will  measure the $(e^++ e^-)$ spectrum from 10 GeV to 20 TeV for the first time directly with fine energy resolution. DM decaying to three leptons may still be a candidate for a DM-only explanation of the positron excess, despite strong constraints from anti-proton and diffuse $\gamma$-ray measurements. We studied CALET's ability to discern the signature of such a DM from a single pulsar source by precise measurement of the $(e^++e^-)$ spectrum. We found that a separation between these two possible explanation of the positron excess will be possible with high probability, especially for DM models which show low $\gamma$-ray emission which are potentially compatible with Fermi-LAT data. These models are characterized by DM mass around 1 TeV and the absence of the $\tau\tau\nu$ channel in the decay. The decay through $ee\nu$ and $\mu\mu\nu$ channels creates a hard drop in the $(e^++e^-)$ spectrum, which can be well identified by CALET.

\appendix

\section{Numerical Calculation of Cosmic Ray Propagation}\label{sec:GALPROPparam}

The publicly available code GALPROP~\cite{galprop} is used to determine the propagation parameters for charged CR in the galaxy, which was used for the propagation of DM decay products. GALPROP solves the CR propagation according to eq.~(\ref{eq:GALPROP}) on a grid in space and momentum numerically, which involves diffusion, diffusive reacceleration and momentum loss during propagation. 
\begin{equation}
\label{eq:GALPROP}
\frac{\partial \psi}{\partial t}= Q(x,p) + \nabla \cdot (D_{xx}\nabla \psi ) + \frac{\partial}{\partial p}p^2D_{pp}\frac{\partial}{\partial p}\frac{1}{p^2}\psi -\frac{\partial}{\partial p}\dot{p}\psi 
\end{equation} 
Here $\psi$ is the density of CR particles per unit momentum, $Q$ is the source term, $D_{xx}$ is the spatial diffusion coefficient and $D_{pp}$ is the diffusion coefficient in momentum space, which describes the reacceleration process. The spatial diffusion coefficient is described as 
\begin{equation}
D_{xx}=\beta D_0\left(\frac{R}{R _0}\right)^{\delta}
\label{eq:diffusionindex}
\end{equation} 
where $D_0$ is the normalization constant, $\beta=\frac{v}{c}$ is the ratio of speed of CR particles with respect to the speed of light, $R=\frac{p}{Ze}$ is the rigidity of the particle and $\delta$ is the power law index of the spatial diffusion coefficient~\cite{GALPROP_manual} as a function of rigidity. The diffusion coefficient in momentum space $D_{pp}$ is related to the spatial diffusion coefficient as 
\begin{equation}
 \label{eq:reaccdiff}
D_{pp}D_{xx} = \frac{4p^2 v_A^2}{3\delta (4-\delta ^2)(4-\delta)}
\end{equation}  
where $v_A$ is the Alfven Speed. \par
\begin{figure}[H]
 
 \centering
 \includegraphics[width=1.0\textwidth]{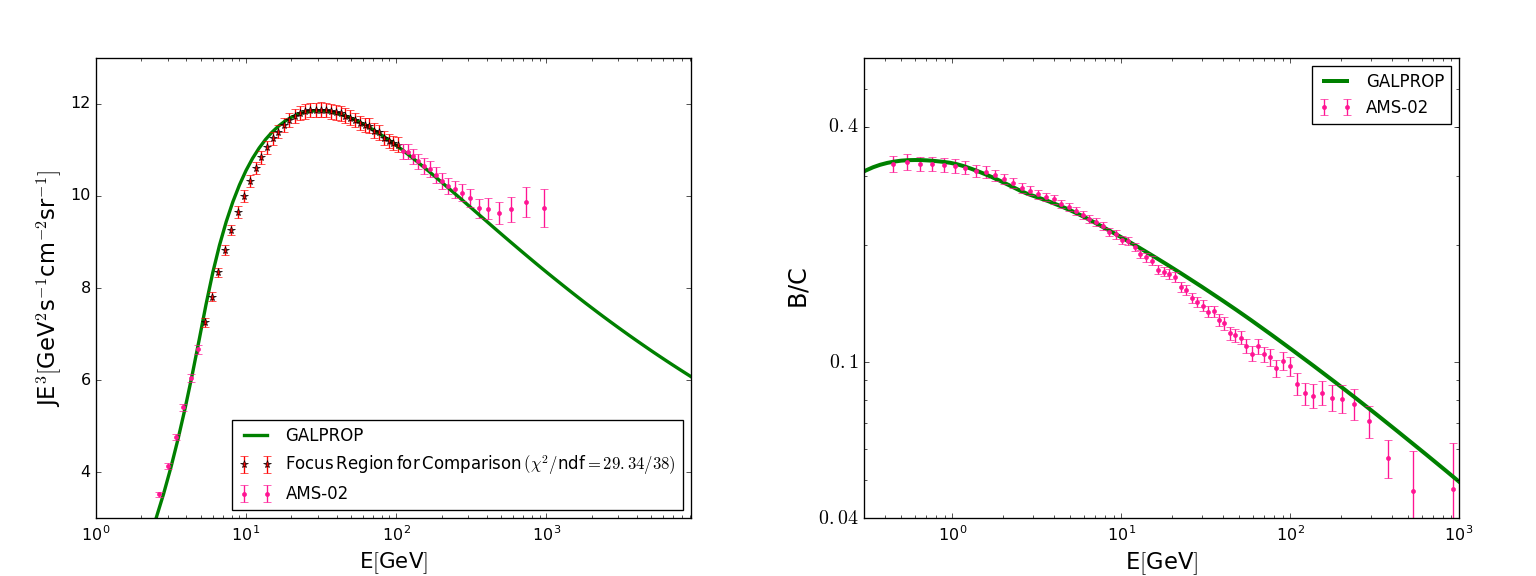}
 \caption{Proton spectrum and $B/C$ ratio calculated with GALPROP (green line) for the propagation parameters given in table~{\ref{Table1}} are compared with the experimental measurements by AMS-02 (magenta dots).}
 \label{fig:GALPROPbkg}
 \end{figure}

A wide range of values for the GALPROP propagation parameters were tested against the  recent results of proton spectra and B/C ratio by AMS-02~\cite{Aguilar:2015ooa, Aguilar:2016vqr} and we concluded on the values given in table~\ref{Table1}. Assuming force field approximation for the solar modulation~\cite{1973Ap&SS..25..387G}, we choose the value of $500$ MeV for the potential~\cite{Cholis:2015gna}. We focus on the range 5\textendash 100 GeV for the comparison with the proton flux measurement since the spectrum above 100 GV progressively hardens as reported by AMS-02~\cite{Aguilar:2015ooa}. The same propagation parameters and solar modulation potential are used for the propagation of heavier nuclei (boron, carbon), electron, positron and also the decay products of the DM. We introduce a low energy spectral break of primary particles where the break is set at a rigidity of \mbox{$7$~GV \cite{Malkov:2010ri}}. As shown in figure~\ref{fig:GALPROPbkg}, the nuclei spectra (proton, B/C ratio) propagated with GALPROP propagation parameters listed in table~\ref{Table1}, are compared with the AMS-02 measurements.

\begin{table}[h]

 \begin{center}
	\begin{tabular}{ | l | l | l |}
	\hline
	\textit{Parameter} & \textit{Value} & \textit{Unit} \\ \hline
	Z$_{\text{max}}$/$\Delta \text{Z}$ & $6/0.25$ & kpc \\ \hline
	X$_{\text{max}}$/$\Delta \text{X}$ & $16/0.25$ & kpc \\ \hline
	Y$_{\text{max}}$/$\Delta \text{Y}$ & $16/0.25$ & kpc  \\ \hline
	$E_{\text{min}}$ & $10$ & MeV \\ \hline
	$E_{\text{max}}$ & $100$ & TeV \\ \hline
	$D_{0}$ (Diff. coeff.)  & $2.90\times 10^{28}$ & $\text{cm}^2\,\text{s}^{-1}$ \\ \hline
	$R_0$ (ref. rigidity for diff. coeff.) & $4$ & GV \\ \hline 
	$\gamma _1/\gamma _2$ (injection index) & $1.70/2.45$ & \\ \hline
        $R_{\gamma}$ (Break in injection Index) & $7$ & GV \\ \hline
	$\delta$ (Diff. coeff. index) & $0.40$ & \\ \hline 
	$v_A$ (Alfven Velocity) & $12.0$ & $\text{km}\,\text{s}^{-1}$  \\ \hline 
	start-timestep & $6.4\times 10^7$ & years \\ \hline 
	end-timestep & $10$ & years \\ \hline
	timestep-factor & $0.90$ & \\ \hline 
	timestep-repeat & $20$ & \\ \hline 
	
	\end{tabular}
	
 \end{center}	
 \caption{GALDEF file parameters used for CR propagation in GALPROP.\label{Table1}}
\end{table}

\par
\begin{figure}[H]
 
 \includegraphics[width=1.0\textwidth]{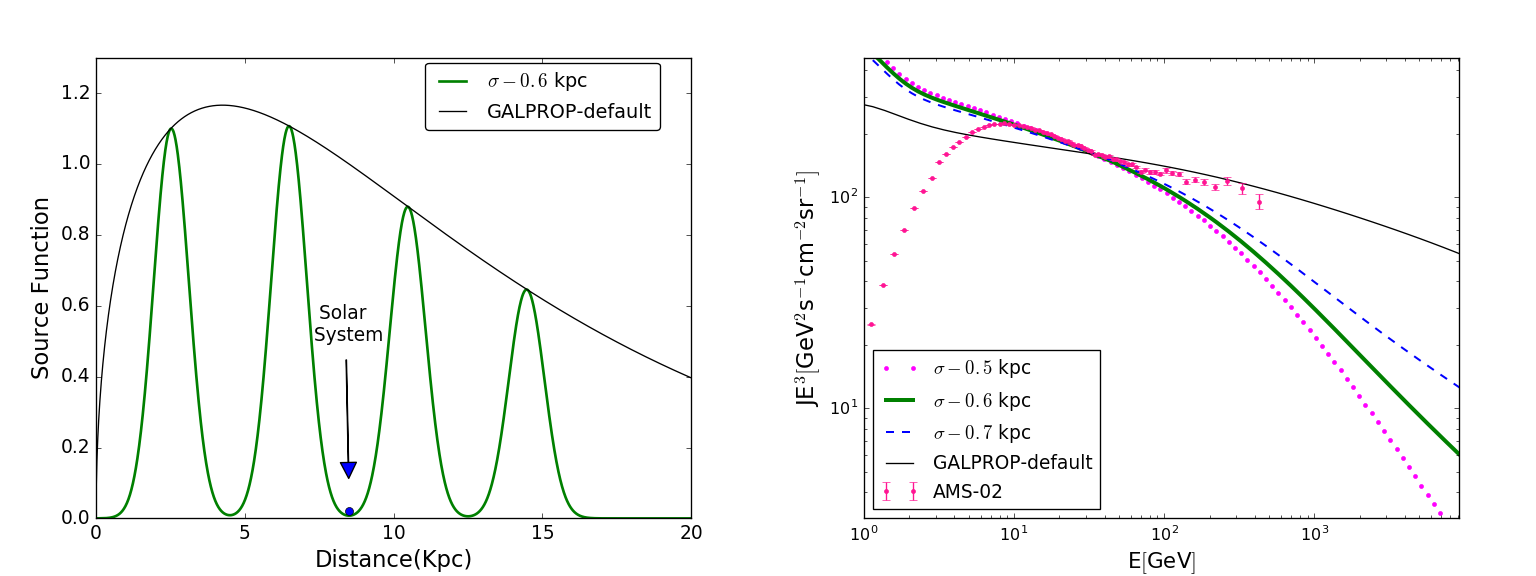}

 \caption{ In the left panel we show the modified source function (in green thick line) with $\sigma = 0.6$ kpc (eq. (\ref{eq:newdis})) compared with the original GALPROP source function (black thin line). Position of the solar system is shown with the blue dot. Dependence of the $(e^++e^-)$ spectrum on $\sigma$ is shown in the right panel.\label{fig:GALPROPsigma}}
\end{figure}

 It is shown in figure~\ref{fig:GALPROPsigma}, that using the same propagation parameters and spectral indices for nuclei and electrons, the electron spectrum obtained from GALPROP is too hard to match the AMS-02 observation at all, even without the addition of an extra source needed for explanation of the positron excess. The GALPROP source distribution is modeled after the SNR distribution derived from the EGRET $\gamma$-ray observation~\cite{Thompson:2008rw}. The spatial distribution of the source function in GALPROP~\cite{Strong:1998pw} is defined as  
\begin{equation}
 \label{eq:sourcedisGAL}
q= q_0 \left(\frac{d}{d_0}\right)^{\eta}\text{exp}\left(-\zeta \frac{d-d_0}{d_0}-\frac{\left|Z\right|}{0.2}\right) 
\end{equation}
where $q_0$ is the normalization constant, and $\eta ,\,\zeta$ are taken as $0.5$ and 1 respectively. In 3D propagation $d$ is taken as $\sqrt{X^2+Y^2}$ and $d_0$ is the distance of the solar system from the center of the galaxy, set to $8.5$ kpc. 
However to represent the spiral arm structure of our galaxy ~\cite{1538-3873-121-877-213,0004-637X-722-2-1460,Gaggero:2013rya,Werner:2014sya}, the source distribution is modeled as 4 concentric rings rings with a Gaussian density profile assuming a half-width $(\sigma)$ in the range $0.5$\textendash $0.7$~kpc. This new spatial distribution of the source function is given by 
\begin{equation}
 \label{eq:newdis}
q_N = q\times \left( \sum _{i=1}^4 e^{-\frac{(d-r_i)}{2\sigma ^2}} \right)
\end{equation}
here $r_i$ are the distances of the ring profile centers from the center of the galaxy. Compared to the original GALPROP source distribution, this spiral arm structure causes the primary cosmic rays to propagate on average a larger distance and experience more energy loss, which makes the CR electron spectra softer. A comparison between the new source distribution and the GALPROP source distribution is shown in the left panel of figure~\ref{fig:GALPROPsigma}.  The effect of the thickness of the spiral arms, which is represented by the $\sigma$ parameter in eq.~(\ref{eq:newdis}), on the $(e^++e^-)$ spectrum is shown in the right panel of figure~\ref{fig:GALPROPsigma}. \par Since the nuclei, electrons and positrons are propagated in GALPROP in one run, we used a high value for the timestep-factor $(0.90)$ and 10~years for the end-timestep~\cite{GALPROP_manual}.

\begin{figure}[h] 
\centering
	\begin{subfigure}[b]{0.92\textwidth}
 	\includegraphics[width=1.0\linewidth]{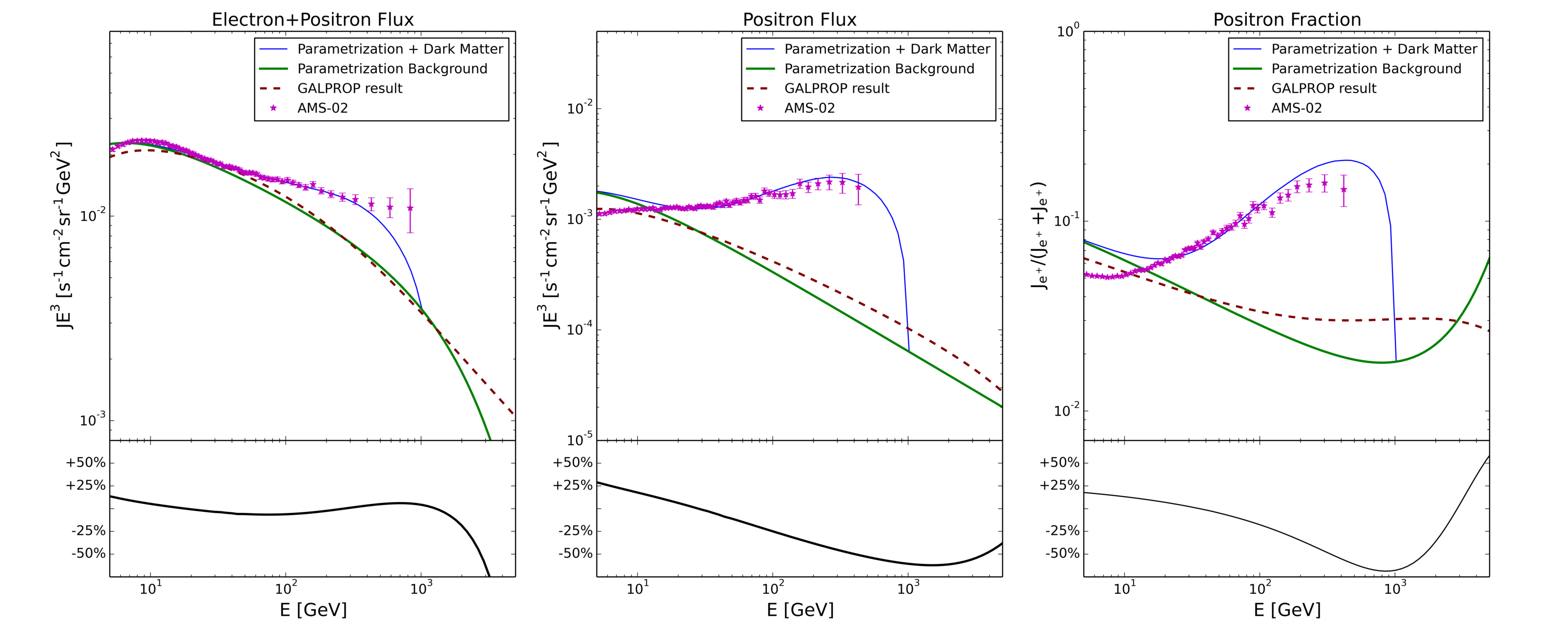}
 	\caption{\label{fig:comparagalnovela}}
\end{subfigure}
	\begin{subfigure}[b]{0.92\textwidth} 
	\includegraphics[width=1.0\linewidth]{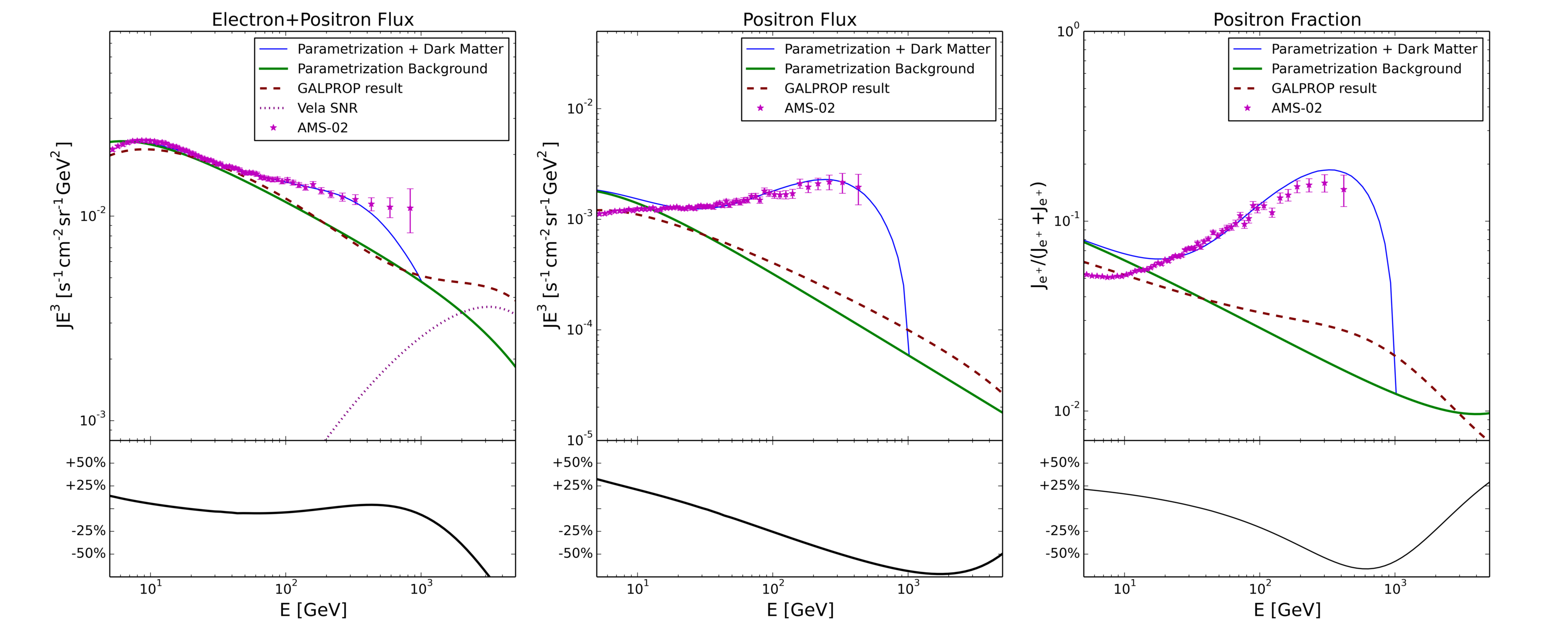}
	\caption{\label{fig:comparagalvela}}
\end{subfigure}
\caption{(a) CR spectra calculated with GALPROP (red dotted line) with $\sigma$ set to \mbox{0.6 kpc} are compared with the parametrization (green line) with DM as extra source and a background cut-off energy $(E_d)$ of 2~TeV. In the lower panel the fractional difference between the GALPROP results and parametrization are shown. (b) Same as figure~\ref{fig:comparagalnovela} but now with $\sigma$ set to 0.5~kpc for the GALPROP calculation and $E_d = 10$~TeV in the parametrization to which Vela SNR (shown with magenta dots) is added. \label{fig:paraGAL}}
\end{figure}

To link the fitted values of the parameters of the parametrization described in section~\ref{sec:Para}, with a specific model of CR propagation, parametrized flux is compared with the GALPROP propagation calculations and the results are shown in \mbox{figure~\ref{fig:paraGAL}}. In the lower panel of \mbox{figure \ref{fig:paraGAL}}, we show the deviation between GALPROP calculation and parametrization and in the relevant energy range the difference is on the order of $70\%$ at most. As shown in figure~\ref{fig:comparagalnovela}, the GALPROP calculation with $\sigma$ set to $0.6$~kpc corresponds to a  value of 2~TeV for the background energy cut-off $(E_d)$ in the parametrization. Variation of the background cut-off parameter $(E_d)$ in the parametrization represents different values of the $\sigma$ parameter in the numerical calculation which represents the ring thickness in this new GALPROP source distribution. \par 

Electron-only flux from Vela SNR is calculated in GALPROP assuming propagation parameters as listed in table~\ref{Table1} except that the time progression is taken as 1200 steps of 10~years and the spatial grid distance is 0.1~kpc in a cube of 12~kpc calculated on the solar system. This flux added to the GALPROP calculated spectrum from distant SNR with $\sigma =0.5$~kpc corresponds best to the parametrization with a value of 10~TeV for the background energy cut-off parameter $E_d$ as is shown in figure~\ref{fig:comparagalvela}. Emission of CRs from the Vela SNR is assumed to be instantaneous and the total energy emitted as electron above 1~GeV normalized to $10^{48}$~erg~\cite{Kobayashi:2003kp}. 

\acknowledgments

We would like to thank Dr. K.Kohri for valuable discussions about calculating the 3-particle decay spectra. S.B. is supported by JICA scholarship. Hardware bought from \textit{FY2016 Waseda University Grant for Special Research Projects Category B, Project No. 2016B-144} (research representative H.M.) was used for CR propagation calculations.


\bibliographystyle{JHEP}
\bibliography{ref}







\end{document}